\documentclass[twocolumn,showpacs,amsmath,amssymb,prl]{revtex4}

\usepackage{epsfig}
\topmargin
-0.2in

\begin{document}
\title{Effective potentials for Folding Proteins}

\author{Nan-Yow Chen$^{1}$, Zheng-Yao Su$^{2,3}$, Chung-Yu Mou$^{2,4}$}
\affiliation{
1. Institute of Physics, Academic Sinica, Nankang, Taiwan\\
2.Department of Physics, National Tsing Hua University, Hsinchu, Taiwan \\
3. National Center for High-Performance
Computing, Hsinchu, Taiwan\\
4. National Center for Theoretical Sciences, Hsinchu, Taiwan}
\date{\today}

\begin{abstract}
A coarse-grained off-lattice model that is not biased in any way to
the native state is proposed to fold proteins.
To predict the native
structure in a reasonable time, the model has included
the essential effects of water in an effective potential.
Two new ingredients, the dipole-dipole interaction and the local
hydrophobic interaction, are introduced and are shown to
be as crucial as the hydrogen bonding.
The model allows successful
folding of the wild-type sequence of protein G
and may have provided important hints to
the study of protein folding.
\end{abstract}

\pacs{87.15.Aa, 36.20.-r, 87.14.Ee}

\maketitle
 The problem of predicting the native structure of a protein
 for a given sequence has been of great interest due to its
 relevancy to many fields in biology.
 In the crudest level, lattice models are proposed and
 have provided important insights \cite{Abkevich,KADill}; however,
 due to the oversimplification,
 they are far from real applications.
 On the other hand, all-atom simulations
 deliver more details for the folding process,
 but the requirement of computational resources tends to
 be realistically unaffordable \cite{Duan2}.
 Developing models of coarse graining thus becomes the
 next step.
 For this purpose, off-lattice models \cite{JNOnuchic1}
 using G$\bar{o}$-type \cite{NGo} potentials have been
 used to explore the folding dynamics. Since the relevant interactions
 are based on native structures, the
 G$\bar{o}$-type potentials can not be used to predict structures.
 There are also models that succeeded
 in separately folding helix bundles
 or folding beta hairpins \cite{Thirumalai2}.
 Nevertheless, the interacting potentials
 employed are also biased towards the native states.
 So far, there is no model that can fold proteins using realistic
 potentials, it is therefore desirable to construct a coarse-grained model
 that can fold proteins without being biased in any way to the native state.

 In this paper, based on microscopic considerations,
 we propose a coarse-grained model with
 realistic potentials.
 The model has been tested successfully on more than 16
 small proteins, of sizes from 12 to 56 amino acids \cite{protein}.
 For most examples,
 even without particularly optimizing our code,
 the computing time is reasonably short and is within
 the order of hours on ordinary desktop computers.
 Here, instead of exploring its predicting ability,
 we shall be focusing on only one protein
 (one of the protein G families with PDB ID : 1GB4) to illustrate
 the folding mechanism embedded in the proposed model. A brief summary
 of other important proteins is given in \cite{protein}.

 In our model, side-chains are coarse-grained as spheres but
 explicit structures are kept in backbones \cite{data}.
 On the other hand, water molecules are not included explicitly
 but their effects are incorporated in effective potentials
 among side-chains and backbones.
% To devise these potentials, it is important to realize that
% they are either indirect effects due to water or originate from
% molecular interactions.
 The hydrophobic (HP) interaction
 has been known as the most important effect due to water.
 Recently, it is realized that the length-scale of water
 molecules
 has to be kept at short distances to prevent proteins collapsing
 prematurely \cite{Humer2}. Therefore, the desolvation model \cite{Humer2}
 combined with the Miyazawa-Jernigan (MJ) matrix \cite{MJ}
 is employed to describe the interaction among the side chains.
 Furthermore,
 since the MJ potential is a non-neighboring interaction,
 its extension to include nearest neighbors (n.n.) along
 the sequence is needed.
 Similar to the spirit of the HP model \cite{KADill},
 a {\em local hydrophobic potential}, $V_{LocalHP}$,
 is implemented by assigning potential energies
 to any successive pairs of amino acids
 according to their hydrophobicity.
 On the other hand,
 the hydrogen bonding (HB) has long been thought as
 the key molecular interaction \cite{Thirumalai2}.
 However, for small proteins, it is known that HB prefers
 the helix structure over the beta sheet because
 the former has a larger number of HBs.
 Thus it hints to include a second molecular interaction.
 Indeed, analysis on the MJ matrix indicates that
 the electric dipole-dipole interaction
 dominates in the pair-wise interaction among side chains \cite{HCLee}.
 Microscopically,
 there is also charge imbalance in the CO-NH group on the amide plane
 with the magnitude of the dipole being estimated to be
 $p=1.15 \times 10^{-19}Cm$.
 Simple analyses reveal that {\em the directions of these dipoles have
 strong correlation with the secondary structure} \cite{dipole}: In the
 alpha helix, successive dipoles on the backbone
 tend to be in parallel; while in the
 beta sheet, they tend to change directions alternately(see Fig.~1 for
 example).
 In order to capture relevant energetics,
 we explicitly introduce
 the {\em dipole-dipole interaction} $V_{DD}$ among the backbone elements.
 The potentials $V_{LocalHP}$ and
 $V_{DD}$ are the main ingredients
 that make our model different from early models. Remarkably,
 our simulations indicate that these two interactions and
 the hydrogen bonding form the key interactions for
 determining the secondary structure. Specifically,
 we find that while the hydrogen bonding is essential to the formation of
 the alpha helix, to fold the beta sheet,
 both $V_{DD}$ and $V_{LocalHP}$ are indispensable.

The potential is constructed
in a renormalized fashion: Except for global multiplicative scales (denoted
by $\epsilon_{\alpha}$ in the following,
with $\alpha$ representing different contributions),
interactions (such as $V_{DD}$ and $V_{MJ}$, see below)
at large distances take the usual form; while for
interactions (such as $V_{LocalHP}$ and $V_{ND}$, see below)
at successive neighbors (short distances),
since the variation of distance is unimportant, only
angle variables are kept.
The parameters employed in the potentials
are adopted from experimental data \cite{data}, while
the scales $\epsilon_{\alpha}$'s are calibrated based on a few proteins
of known structures \cite{calibration}.
% For long-range molecular interactions,
% we take the same microscopic form at large
% distances but leave global renormalization
% scales, due to the effect of water molecules, to
% be fixed by experimental data.
% The renormalization ceases
% to be valid for short distances. In this case,
% they are replaced
% by nearest-neighbor interactions.
% On the other hand, if an interaction is only known at
% short distances such as the MJ matrix, analytical continuation
% to large distances is performed with an undetermined global scale.
% These global scales (denoted by $\epsilon$)
% are calibrated based on a few proteins of known structures\cite{calibration}.

 The degrees of freedom for backbones
 are two Ramachandran angles
 $\phi$ and $\psi$ \cite{Ramachandran}.
 Since the peptide bond on any amide plane
 is partially double-bonded, the angle $\omega$ around the
 peptide bond is fixed to be $180^{\circ}$ so that it
 corresponds to the {\em trans} conformation.
 The spheres that represent side-chains are centered at $C^{\beta}$ and
 are attached to  $C^{\alpha}$-atoms rigidly, and
 different effective radii are assigned in consistent with the
 geometric structures \cite{radius}. In these representations and
 with all energies being in unit of kcal/mol,
 the potential can be written as
 \begin{equation}
 V_{total}=V_{steric}+V_{HB}+V_{DD}+V_{MJ}+V_{LocalHP}+V_{A}.
 \label{Vtotal}
 \end{equation}
 Here $V_{steric}$ enforces structural constraints such as hard-core
 potentials to avoid unphysical contacts. $V_{HB}$ accounts for
 the hydrogen bonding between any non-neighboring $NH$ (labeled by $i$) and
 $CO$ ($j$) pair and is implemented as
% \begin{equation}
%V_{HB}=\epsilon_{HB} \sum_{n,i,j}u\left( r_{ij} \right) v \left(
%\theta_{n,ij} \right), \label{VHB}
%\end{equation} where
$ V_{HB}=\epsilon_{HB} \sum_{n,i,j} u( r_{ij} ) v (
\theta_{n,ij} )$,
where $r_{ij}$ is the distance between $H_{i}$ and
$O_{j}$ and
$u(r)$ is the standard 12-10 Lennard-Jones potential
with the equilibrium distance being set to the
the averaged experimental value $1.738 \AA$ \cite{data}.
The angle function $v$ imposes the directional nature of HB, parameterized
by three angles ($n=1,2,3$):
$\pi -\angle C_{i}O_{i}H_{j}$,
$C_{i}O_{i} \wedge N_{j}H_{j}$, and $\pi - \angle O_{i}H_{j}N_{j}$.
Their values are confined to the averaged experimental data \cite{data}
respectively: $26.77^{\circ}$, $11.60^{\circ}$,
and $17.98^{\circ}$.
To increase the efficiency of HB formation,
certain uncertainty $\Delta \theta$ is allowed.
Empirically, $\Delta \theta =60^{\circ}$ is most efficient.

The dipole term $V_{DD}$
at large distances takes the ordinary form
\begin{equation}
V_{DG}=\epsilon_{DG}\sum_{i,j} \left[  \frac{\vec{p}_{i} \cdot \vec{p}_{j}
}{r_{ij}^{3}}-\frac{3 \left( \vec{p}_{i} \cdot \vec{r}_{ij}\right)
\left( \vec{p}_{j}
\cdot \vec{r}_{ij} \right) }{r_{ij}^{5}} \right], \label{VDD}
\end{equation} where $\vec{p}_{i}$ and $\vec{p}_{j}$ are dipoles of
either $CO$ or $NH$, and the
summation excludes successive dipoles.
When dipoles are in successive neighbors, it is
given by
\begin{equation}
V_{DN}=\epsilon_{DN}\sum_i \frac{1}{2} \left( \frac{\vec{p}_{i}
\cdot \vec{p}_{i+1}}{ {p_{i}}  {p_{i+1}} }-1
\right). \label{VDN}
\end{equation}
$V_{MJ}$ is the extension of the MJ matrix with the form
$ V_{MJ}  = \epsilon_{MJ}\sum_{i,j} [
V_{LJ} \left( r_{ij} \right) + V_{G1} \left( r_{ij} \right) +
V_{G2} \left( r_{ij} \right) ]$.
%\begin{equation}
%V_{MJ}  = \epsilon_{MJ}\sum_{i,j} \left[
%V_{LJ} \left( r_{ij} \right) + V_{G1} \left( r_{ij} \right) +
%V_{G2} \left( r_{ij} \right) \right]. \label{VMJ1}
%\end{equation}
Here $V_{LJ}$ is the MJ matrix element $\epsilon_{ij}$ multiplied by
the usual 12-6 Lennard-Jones potential
with the equilibrium distance being the sum of radii of two side-chains.
$V_{G1} + V_{G2}$ represents the potential obtained numerically
in the desolvation model \cite{Humer2}.
For numerical purpose, however,
we find that it is more convenient
to use the following approximately
analytic forms:
$ V_{G1} = \epsilon_{1} \times \exp \left[
-\sigma _{w} \times \left( r_{ij}-r_{b} \right) ^{2} \right] $
is a Gaussian fit to the desolvation barrier
with $r_{b}$ being the position of
desolvation barrier and $\sigma _{w}$ being the
size of the water molecule;
 while
$ V_{G2} = \epsilon_{2} \times \exp \left[
-\sigma _{w} \times \left( r_{ij}-r_{w} \right) ^{2} \right]$
is an inverted Gaussian fit to the metastable minimum
at $r_{w}$ due to water molecules.
Here for the best fit, $\epsilon_{1}
\sim 5 |\epsilon_{ij}| /9 $ and
$\epsilon_{2}
\sim - |\epsilon_{ij}| /3 $.

The potential $V_{LocalHP}$ acts only on successive pairs of side-chains
\begin{equation}
V_{LocalHP}  = \sum_{i}
V_{q_i,q_{i+1}} .
\label{LocalHP}
\end{equation}
Here $q_i$ represents the hydrophobicity or the charge state
of the $i$th side-chain. Following Ref.\cite{dipole} ,
$q_i$ are classified
into hydrophobic(H),
polar(P), neutral(N), positive charged($+$), and
negative charged($-$). In this classification, N is regarded as a referential
type such that whenever $q_i=N$ or $q_{i+1}=N$ ,
$V_{q_i,q_{i+1}}=0$. Furthermore, when
charged side-chains encounter other non-charged ones, they are considered
as polar. Therefore, the only nontrivial potential energies are
$(V_{HH}, V_{PP}, V_{+-})$ (attractive) and $(V_{HP}, V_{++})$
(repulsive). To implement the hydrophobic effects, an attractive
pair acquires a negative energy $-\epsilon_{q_i,q_{i+1}}$
when their $C^{\alpha} C^{\beta}$ lines
are parallel to each other, and when in other orientation,
no energy is assigned;
while for repulsive pairs, a negative energy $-\epsilon_{q_i,q_{i+1}}$
is assigned
when their $C^{\alpha} C^{\beta}$ lines are anti-parallel. In practice,
a smooth function is used to interpolate between
finite $V_{q_i,q_{i+1}}$ and zero.
Finally, $V_{A}$ is an on-site potential in proportion to
the area of each side-chain that is exposed to water.
The proportional constant is $\epsilon_{ii}- \langle \epsilon_{ii}
\rangle $ with $i$ being the index for the side-chain.
The existence of $V_A$ has already been found in
the analysis of the MJ matrix \cite{HCLee} and it helps to further
contrast the hydrophobicity of each side-chain.

The Monte Carlo method is employed to fold proteins.
After careful calibration \cite{calibration}, the global scales are found to
be $\epsilon_{HB} \approx 4.8$, $\epsilon_{DG} \approx 0.2$,
$\epsilon_{DN} \approx 2.1$, $\epsilon_{MJ} \approx 0.2$,
and for $V_{LocalHP}$, $\epsilon_{HH}=\epsilon_{PP}=\epsilon_{HP} \approx 5.0$,
$\epsilon_{++} = \epsilon_{+-} \approx 5.0$. The same scales are adopted
to simulate the protein 1GB4, which
is a wild-type protein with
one alpha helix and two beta hairpins.
Fig.~1 shows
its spatial arrangement and corresponding
dipole arrangement of our simulated energy ground state, while
Fig.~2 shows the contact map.  The native contact number ratio
($Q$) for simulated ground state is $0.6$, while the RMSD
is 2.97 $\AA$.
Clearly, our simulation is in good agreement
with the experiment while the computing time is only a
few hours on a P4-3.0GHz PC.  Note that the ground state
energy is -545 kcal/mol and the nearest local minimum
is about 34 kcal/mol higher in energy. Furthermore,
both the helix and the beta sheet are formed
only when correct scales $\epsilon $'s and
appropriate temperature are employed. The portability of
these scales (and our model) to other proteins are tested in 15 proteins.
The results are briefly summarized in Ref.\cite{protein}.
Our results are generally
in good agreement with experiments with the tolerance of
$\epsilon $'s being about $0.5$. Occasionally,
the accuracy is not good. However, in that case, the cause
is due to the metal ion not being included in our simulation \cite{protein}.
\begin{figure}
\hspace*{-10mm}
\includegraphics*[width=80mm]{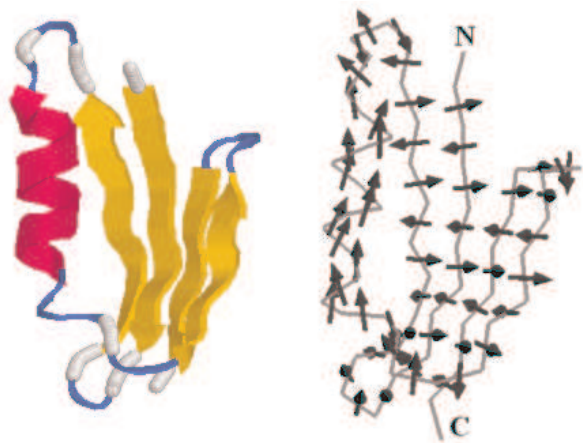}
\caption{\small Schematic plot of the simulated protein G - 1GB4:
native conformation and corresponding dipole configuration.
}
\end{figure}
\begin{figure}
\hspace*{-12mm}
%\rotatebox{-90}{\includegraphics*[width=70mm]{fig2new2.eps}}
\includegraphics*[width=70mm]{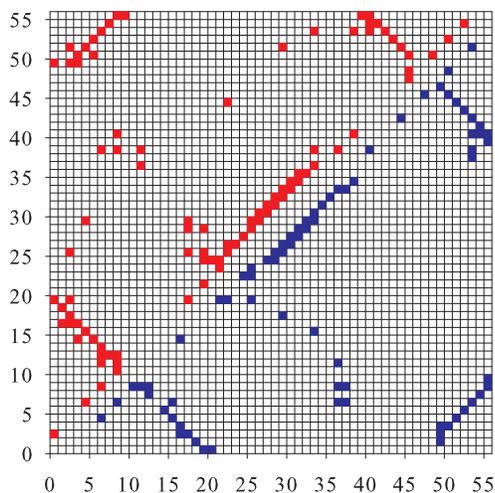}
\caption{\small Contact map of energy ground states (native
structures) for 1GB4 at $k_B T=0.8$ kcal/mol with blue square
being our simulation result ($Q=0.6$) and red square being the
data from PDB. } \label{contact}
\end{figure}

To clarify the roles of $V_{DD}$ and
$V_{LocalHP}$, the alpha helix (A24 to D37)
and the beta hairpin with C terminus (G42 to E57)
are extracted.
The energy versus $Q$
along the folding
is then monitored for different strengths of the potentials.
Figs.~3(a) and 3(b) show the effect
of $V_{DD}$ for three different strengths.
Clearly, we see that the native conformation ($Q=1$)
stays at the minimum for the helix, while for the beta hairpin,
it gradually moves away from the minimum.
When $V_{DD}$ is completely turned off, the beta sheet is no longer the
ground state.
\begin{figure}
\hspace*{0mm}
%\rotatebox{-90}
{\includegraphics*[width=80mm, height=64mm]{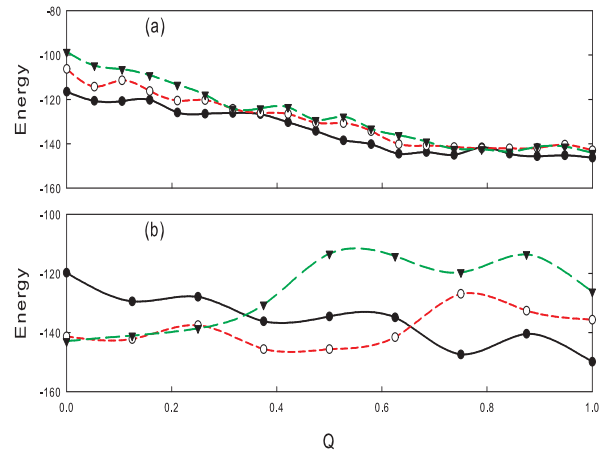}}
\caption{\small Effects of different strengths of $V_{DD}$ on the
formation of the alpha helix $(a)$ and the beta hairpin with C
terminus  $(b)$. The corresponding strengths: solid (black) -
$V_{DD}$, circle (red) - $0.5V_{DD}$, and triangle (green) - $0$.
$Q$ is the native contact number ratio with $Q=1$ corresponding to
the native conformation.} \label{fig3}
\end{figure}
Similar analyses are done by tuning $V_{LocalHP}$
as shown in Figs.~4(a) and 4(b).
We see that although affecting the formation of the alpha
helix, both $V_{LocalHP}$ and $V_{DD}$
have stronger effects on
the formation of the beta sheet and can change
the ground state completely. Similar behaviors also occur
for the beta hairpin with N terminus and other 15 proteins.
For 1GB4,
if we turn off $V_{LocalHP}$ and $V_{DD}$, only segments
of helices are formed.
Therefore,
both $V_{DD}$ and $V_{LocalHP}$ are responsible for
the formation of the beta sheet.
\begin{figure}
\hspace*{0mm}
%\rotatebox{-90}
{\includegraphics*[width=80mm, height=64mm]{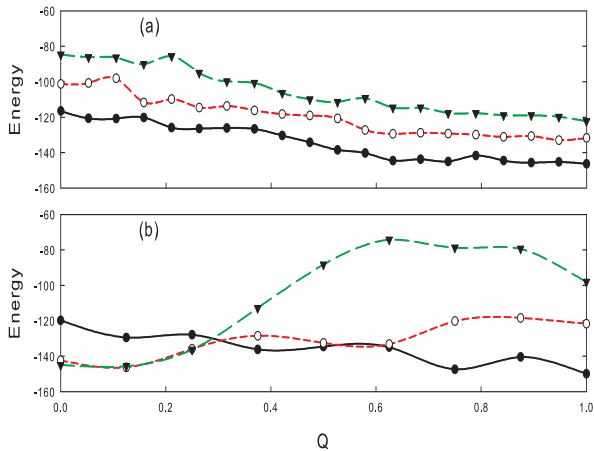}}
%\includegraphics*[width=80mm]{Fig4nan.eps}
%\vspace{-5mm} \\
%\hspace*{10mm}
%\rotatebox{-90}{\includegraphics*[width=60mm]{fig4b.eps}}
\caption{\small Effects of different strengths of $V_{LocalHP}$ on
the formation of the alpha helix $(a)$ and the beta hairpin with C
terminus $(b)$. The corresponding strengths: solid (black) -
$V_{LocalHP}$, circle (red) - $0.5V_{LocalHP}$, and triangle
(green) - $0$. } \label{fig4}
\end{figure}

It should be noted that the above analyses are done with
fixed $V_{HB}$, and the vanishing alpha helix in Fig.~4(a)
can be restablized by increasing $V_{HB}$.
However, similar restablization
does not occur to the beta sheet due to the fact
that the helix has more HBs. Therefore, when
$V_{HB}$ is large enough, the helix conformation always
wins, and even a beta sheet will be turned into a helix.
On the other hand, because successive dipoles in a helix
tend to have unfavorable parallel orientations,
sufficient strong $V_{DD}$ can stabilize
the beta sheet over the helix. Therefore,
in the intermediate strength of $V_{HB}$, a beta
sheet could form if
the deficient energy due to smaller number of HBs
is compensated by the energy gain of $V_{DD}$.

Similar analysis on the MJ potential
shows that instead of
deciding the secondary structure explicitly,
$V_{MJ}$ plays a crucial role in making its formation
more efficiently.
In early stage of folding, $V_{MJ}$ collapses
all residues into a compact space.
Only when the collapsing happens, interactions of shorter ranges
could function.
If the initial collapsing does not go in the
right direction or happens too fast,
the final protein structure may become disordered.
After the initial collapsing, the potentials
$V_{DD}$, $V_{LocalHP}$, and $V_{HB}$ start to dominate.
At this point, an obvious question remains to be addressed:
Since both $V_{DD}$ and $V_{HB}$ are sequence independent,
then for a given sequence, what determines that it should
fold into a helix or a beta sheet?
This is where $V_{LocalHP}$ comes into play
because it forces successive neighboring side-chains
to be either on the same side or on the opposite side of the
backbone according to their hydrophobicity.
Thus different sequences result in different
local spatial arrangements of side-chains, and only when the
arrangement is correct, the protein can be compacted
into the correct secondary structure. Finally, our analysis
shows  that even though the native state is still
the ground state in the absence of $V_A$,
incorrect strength of $V_A$ would result in itinerant
motion of the secondary structures. Therefore,
$V_A$ is primarily responsible for stabilizing
the tertiary structure.

In conclusion, an effective potential that can fold proteins
without being biased
to the native state is constructed and tested.
All testing peptides can fold to their native states in
acceptable computing time.
By systematically tuning relative strengths of
interactions in the potential,
the dipole-dipole interaction $V_{DD}$
and the local hydrophobic interaction $V_{LocalHP}$
are shown to be as crucial as the hydrogen bonding $V_{HB}$.
While $V_{HB}$ prefers the helix structures,
$V_{DD}$ tends to form sheet-like structures.
Only when a subtle balance between these
two interactions holds, the helix and sheet structures can
co-exist. The sequence-dependent potential
$V_{LocalHP}$ is then responsible for the final selection of
either a helix or a beta sheet forming.

We thank Profs. C. C. Chang and T. K. Lee
for useful discussions. This research was supported by NSC of Taiwan.

%\begin{references}

\end{document}